\documentclass[aps,prl,superscriptaddress,preprint]{revtex4-1}

	\usepackage{amsmath}
	\usepackage{amssymb} 
	\usepackage{color}
	\usepackage{makeidx}
	\usepackage{amsfonts}
	\usepackage[ansinew]{inputenc}
	\usepackage[usenames,dvipsnames]{pstricks}
	\usepackage{subfigure}
	\usepackage{epsfig}
	\usepackage[colorlinks,hyperindex]{hyperref}
	\usepackage[squaren,Gray]{SIunits}

	\hypersetup
	{
		colorlinks,%
		citecolor=black,%
		linkcolor=black,%
		urlcolor=black,%
	}

\makeindex
\begin{document}

\title{The Universality of Thermal Transport in Amorphous Nanowires at Low Temperatures}

\author{Adib Tavakoli}
\affiliation{Institut N\'EEL, CNRS, 25 avenue des Martyrs, F-38042 Grenoble, France}
\affiliation{Univ. Grenoble Alpes, Inst NEEL, F-38042 Grenoble, France}

\author{Christophe Blanc}
\affiliation{Institut N\'EEL, CNRS, 25 avenue des Martyrs, F-38042 Grenoble, France}
\affiliation{Univ. Grenoble Alpes, Inst NEEL, F-38042 Grenoble, France}

\author{Hossein Ftouni}
\affiliation{Institut N\'EEL, CNRS, 25 avenue des Martyrs, F-38042 Grenoble, France}
\affiliation{Univ. Grenoble Alpes, Inst NEEL, F-38042 Grenoble, France}

\author{Kunal J. Lulla}
\affiliation{Institut N\'EEL, CNRS, 25 avenue des Martyrs, F-38042 Grenoble, France}
\affiliation{Univ. Grenoble Alpes, Inst NEEL, F-38042 Grenoble, France}

\author{Andrew D. Fefferman}
\affiliation{Institut N\'EEL, CNRS, 25 avenue des Martyrs, F-38042 Grenoble, France}
\affiliation{Univ. Grenoble Alpes, Inst NEEL, F-38042 Grenoble, France}

\author{Eddy Collin}
\affiliation{Institut N\'EEL, CNRS, 25 avenue des Martyrs, F-38042 Grenoble, France}
\affiliation{Univ. Grenoble Alpes, Inst NEEL, F-38042 Grenoble, France}

\author{Olivier Bourgeois}
\affiliation{Institut N\'EEL, CNRS, 25 avenue des Martyrs, F-38042 Grenoble, France}
\affiliation{Univ. Grenoble Alpes, Inst NEEL, F-38042 Grenoble, France}

\date{\today}

\begin{abstract}

Thermal transport properties of amorphous materials at low temperatures are governed by the interaction between phonons and localized excitations referred to as tunneling two level systems (TLS). The temperature variation of the thermal conductivity of these amorphous materials is considered as universal and is characterized by a quadratic power law. This is well described by the phenomenological TLS model even though its microscopic explanation is still elusive. Here, by scaling down to the nanometer scale amorphous systems much below the bulk phonon-TLS mean free path, we probed the robustness of that model in restricted geometry systems. Using very sensitive thermal conductance measurements, we demonstrate that the temperature dependence of the thermal conductance of silicon nitride nanostructures remains mostly quadratic independently of the nanowire section. It is not following the cubic power law in temperature as expected in a Casimir-Ziman regime of boundary limited thermal transport. This shows a thermal transport counter intuitively dominated by phonon-TLS interactions and not by phonon-boundary scattering in the nanowires. This could be ascribed to an unexpected high density of TLS on the surfaces which still dominates the phonon diffusion processes at low temperatures and explains why the universal quadratic temperature dependence of thermal conductance still holds for amorphous nanowires.

\end{abstract}


\maketitle

Amorphous materials may have significant dispersion in their chemical compositions or their physical structures at the microscopic level. However, at low temperatures, the behavior of the thermal properties of almost all amorphous materials are thought to be universal \cite{RevPohl2002}.
 These common features include a nearly linear specific heat and a nearly quadratic thermal conductivity in temperature below few Kelvin. As thermal transport is concerned, this universality is not only qualitative but also quantitative, indeed the thermal conductivity of all amorphous materials lies within a factor of twenty in the same order of magnitude called the \textit{glassy range} \cite{PohlPRB1971,Stephens}.
Despite much theoretical efforts, this universality remains poorly understood and its true microscopic origin is still elusive. Nowadays, the most accepted model is based on the presence of tunneling two level systems (TLS) involving tunneling between different equilibrium positions of an atom or group of atoms \cite{Phillips1972,bibi72,Zaitlin1975}. The scatterings of the phonon on these tunneling sites is assumed to be at the origin of the quadratic variation of thermal conductance in temperature. The phonon heat transport is then characterized by the phonon-TLS mean free path (the distance between two inelastic collisions written down MFP) which is on the order of few hundreds of micrometers.

Phillips suggested that TLS are likely to form in materials with an open structure and low coordination regions, and are unlikely in highly dense amorphous systems \cite{Phillips1972}. Recent experiments give indication of the correlation between the low density regions, the presence of nanovoids and the presence of TLS \cite{queen2013,queen2015}. In the opposite case the different experiments based on hydrogenated Si \cite{liupohl}, and ultrastable glasses \cite{liu2014,Viejo2014} have shown a significant reduction of the TLS density and a tendency to be more crystal-like \cite{liu2014,Viejo2014,liu2015}. These results support Phillips original suggestion.
Understanding the origin of these localized excitations (or TLS) is one of the most challenging problem of modern condensed matter physics at low temperature. Indeed, many questions have been raised concerning their existence \cite{leggett1991}, their fundamental origins \cite{leggett2013}, their possible role in the decoherence of quantum entangled states in Josephson quantum bit \cite{Ku,Simmonds}, or their noise producing aspects in superconducting resonator \cite{Jiansong}. Probing the phonon-TLS scattering through the measurement of the phonon-TLS MFP in low dimensional samples (membranes, nanowires) can bring significant new insights for the understanding of thermal transport in amorphous materials at the nanoscale.

On another hand, in a dielectric single crystal far below the Debye temperature, the phonon mean free path is set by the phonon-phonon interaction leading to the well known cubic power law in temperature of the thermal conductivity ($\kappa(T)\propto T^3$). This MFP can be very long at low temperature because the phonon-phonon interactions become less probable. This leads to boundary scattering limited transport called the Casimir-Ziman regime where phonon scattering only appears on the edges of the materials \cite{Casimir1939,Ziman,Berman}. It has been shown recently that, at the nanoscale, the thermal transport in a single crystal silicon nanowires belongs to this regime. Thermal conductance having variation in temperature very close to the expected cubic power law has been found \cite{Heron1,Heron2,HeronPRB,blanc2014}. 

Then, the low temperature thermal transport in amorphous materials (bulk or very thick film) departs strongly from its single crystal counterparts by its universal quadratic thermal conductivity \cite{RevPohl2002}. This quadratic variation is the distinctive picture of glassy materials, the bulk phonon MFP being limited by the phonon-TLS inelastic interactions which lies in the range of $20\micro $m$<\Lambda^{bulk}_{ph-TLS}<200\micro$m, at 1~K \cite{Phillips1972,bibi72,Zaitlin1975,Yu1987,RevPohl2002}. The present experiments, done on glassy systems of restricted geometries, are putting in competition the Casimir-Ziman regime where phonons are essentially scattered by the boundaries (characterized by a thermal conductance cubic in temperature) and the amorphous regime where phonons are scattered by TLS.
The main objective of this work is then to probe phonon transport in spatially confined systems at the nanoscale, i.e. below the characteristic length set by the bulk phonon-TLS MFP in amorphous materials $\Lambda^{bulk}_{ph-TLS}$. This should yield crucial insights on the location and maybe on the origin of the TLS in glasses.   

Here, we carried out very sensitive thermal conductance measurements on silicon nitride nanostructures at low temperatures. The samples have various dimensions from millimeter membrane to micro and nanowires, where the sizes are purposely much smaller than the bulk phonon MFP in amorphous materials  (set by the phonon-TLS interactions). As silicon nitride is known to be a fully amorphous materials, widely used for its exceptional mechanical and thermal properties \cite{fefferman,Unter2010,hellman1,hellman2,Leivo,goldie,zen2014}, it is one of the best materials to study the competition between phonon boundary scattering and phonon-TLS interactions play a significant role in the thermal transport. 

The low temperature thermal properties of Si$_3$N$_4$ have been already studied by different groups without considering the possible contribution of TLS to the phonon scattering \cite{Leivo,anghel,holmes,hoevers}; however, a little later the problem has been raised by two theoretical works \cite{anghel2007,kuhn2007,withington2011,withington2013}. The present experiments will allow the probing of 1-the phonon-TLS interaction down to the nanometer scale and 2-its effect on the power law of the variation of thermal conductance versus temperature. Unexpectedly, as the Casimir-Ziman regime should be observed in nanowires through the $T^3$ behavior of $\kappa(T)$, a robust $T^2$ is observed, showing that even in restricted geometry, the phonon-TLS scattering is still governing the heat transport.

\begin{figure}
\begin{center}
 \includegraphics[width=8cm]{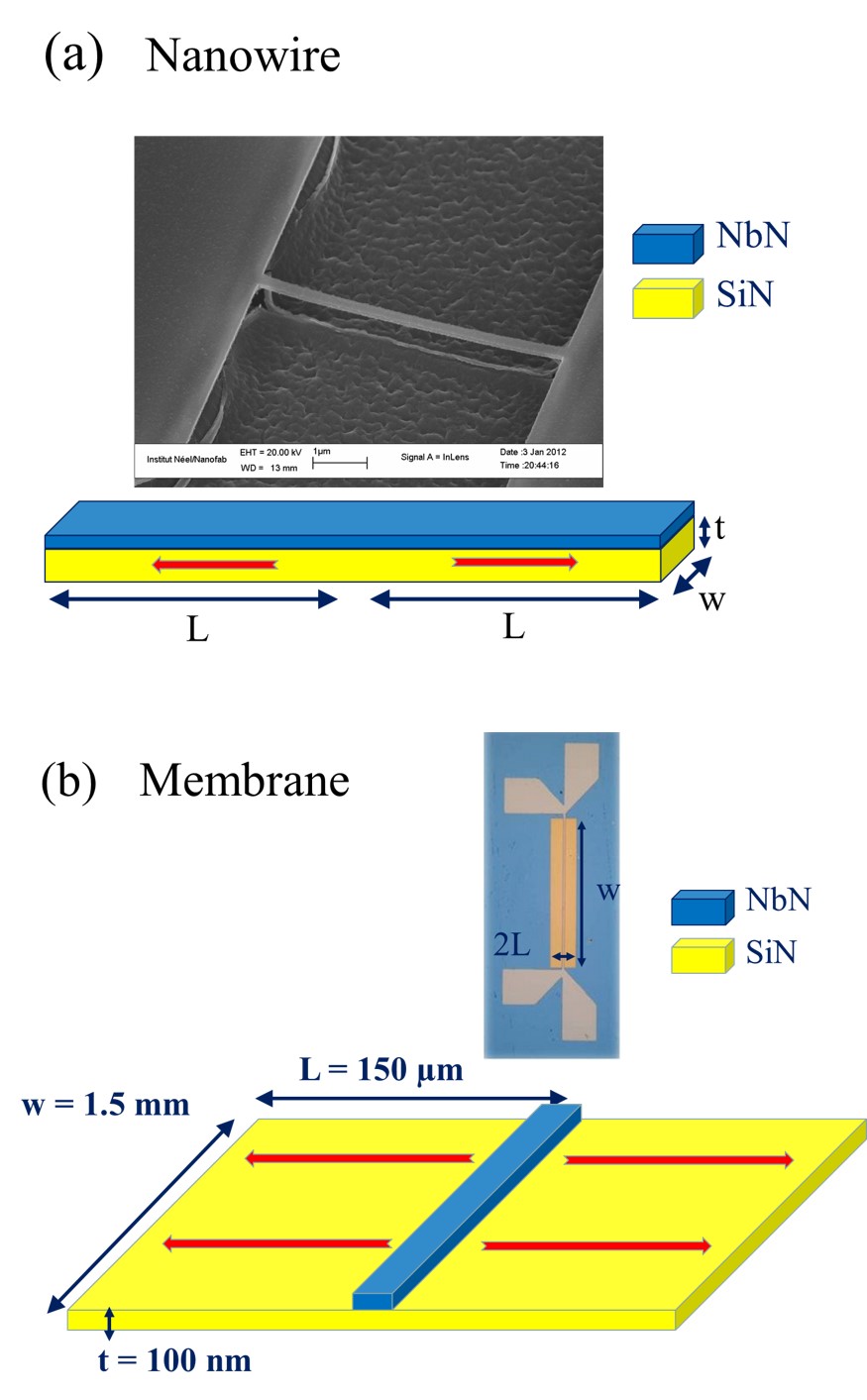}
 \end{center}
 \caption{Pictures and schematic representations of the various Si$_3$N$_4$ suspended structures. The red arrows represent the heat flux, the blue arrows, the dimensions of these samples. (a) Suspended silicon nitride nanowire measured with the longitudinal 3~$\omega$ method \cite{Bourgeois2007} (the microwires are not shown) and (b) experimental configuration of the silicon nitride membrane measured using the planar 3~$\omega$-Volklein method \cite{Sikora2012,Sikora2013}. The blue layer represents the niobium nitride (NbN), the thin film transducer used for the thermal measurements.} 
 \label{fig1}
\end{figure}

The thermal conductance measurements have been performed on 100~nm thick mechanically suspended stoichiometric Si$_3$N$_4$ structures from the millimeter scale (membrane) to the nanometer scale (nanowire) in order to cross the characteristic length given by $\Lambda^{bulk}_{ph-TLS}$ (see Fig.~\ref{fig1}). Various 3$\omega$~methods adapted to each geometry have been used; these different techniques have been already explained elsewhere \cite{Bourgeois2007,Sikora2012,Sikora2013}. All thermal measurements are done using a niobium nitride (NbN) thermometry very sensitive over a broad temperature range (from 0.1~K to 330~K) \cite{thermoNbN}. The thermal conductance of the micro and nanowires are measured using the longitudinal 3$\omega$ technique where the heat flow is along the NbN transducer (see Fig.~\ref{fig1}a). Concerning the membrane, 3$\omega$-Volklein geometry is used, in this technique the heat flow is perpendicular to the transducer since this one is deposited in the center of the membrane (along the long side, see Fig.~\ref{fig1}b). Four geometries of suspended structures have been used for that purpose; all the dimensions of samples are summarized in Table~\ref{table1}. 

\begin{table}
\centering
\begin{tabular}{|c|c|c|c|c|}
  \hline
  Sample type &  w  & L ($\mu$m) & $\zeta_{0}$ (W.K$^{-3}$.m$^{-1})$ & $\Lambda^{1\rm{K}}_{ph}$ ($\mu$m) \\
  \hline
  \hline
membrane &  1.5~mm & 150 & 1.2$\times 10^{-2}$ & 31 \\
	  \hline
large microwire & 7~$\mu$m & 50  & 9$\times 10^{-3}$ & 27 \\
	  \hline
narrow microwire &  1~$\mu$m & 10 & 3$\times 10^{-4}$ &  0.9 \\
	  \hline
nanowire & 200~nm & 2.5 & 1.2$\times 10^{-4}$ & 0.36 \\
	  \hline
	  \hline

\end{tabular}
\caption{Details of the dimensions of the four different kinds of samples made out of Si$_3$N$_4$ thin films: sample types, their width and length (all samples are 100~nm thick). The NbN thermometer is 70~nm thick. The membrane is considered as a semi-infinite sample (very large aspect ratio), and three samples are reduced in dimensions: large and narrow microwires and nanowires. More than three orders of magnitude in sizes are covered by these four samples. The coefficient $\zeta_{0}$  and $\Lambda^{1\rm{K}}_{ph}$, extracted from the thermal conductivity measurements at 1~K, are necessary for the interpretation of the results.}
\label{table1}
\end{table}

\begin{figure}
\begin{center}
 \includegraphics[width=10cm]{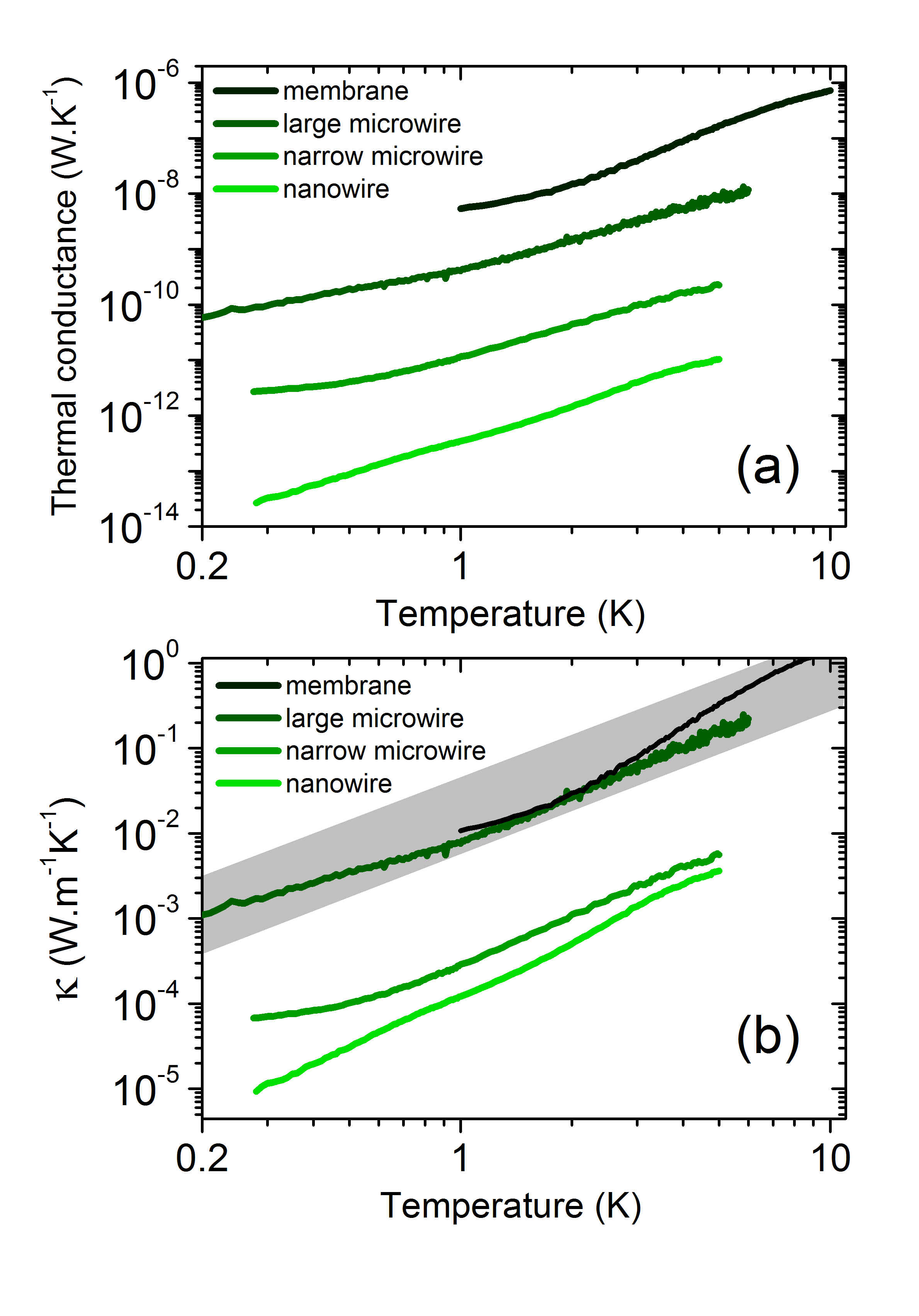}
 \end{center}
 \caption{(a) the thermal conductance of the membrane (black), the large microwire (dark green), the narrow microwire (green), and the nanowire (light green) is shown in a log-log plot. The overall temperature behavior of the thermal conductance of all these samples is quite similar. (b) effective thermal conductivities calculated as a base for comparison of phonon thermal transport between the different geometries. As the dimensions are reduced, the thermal conductivities decrease. The glassy range is represented by the grey area in the plot.}
 \label{fig2}
\end{figure}

In Fig.~\ref{fig2}~(a) the thermal conductance of the nanowire, the microwires and the membrane is shown in a log-log plot. The first point that needs to be highlighted is the similar quadratic temperature behavior for all the different samples with a thermal conductance proportional to $T^{1.5}$ to $T^{2}$. It is in agreement with the universal behavior of glasses $T^{1.8}$, but far from the cubic behavior expected for the Casimir-Ziman regime \cite{RevPohl2002,PohlPRB1971}. 
Quantitatively, the conductance of the nanowire is almost two orders of magnitude below the conductance of the narrow microwire, and six orders of magnitude below the conductance of the membrane. This is the consequence of several concomitant effects: the reduction of the geometry (boundary scattering) and the reduction of the phonon MFP due to phonon-TLS interaction that both limit the heat transport. 

\begin{figure}
\begin{center}
 \includegraphics[width=10cm]{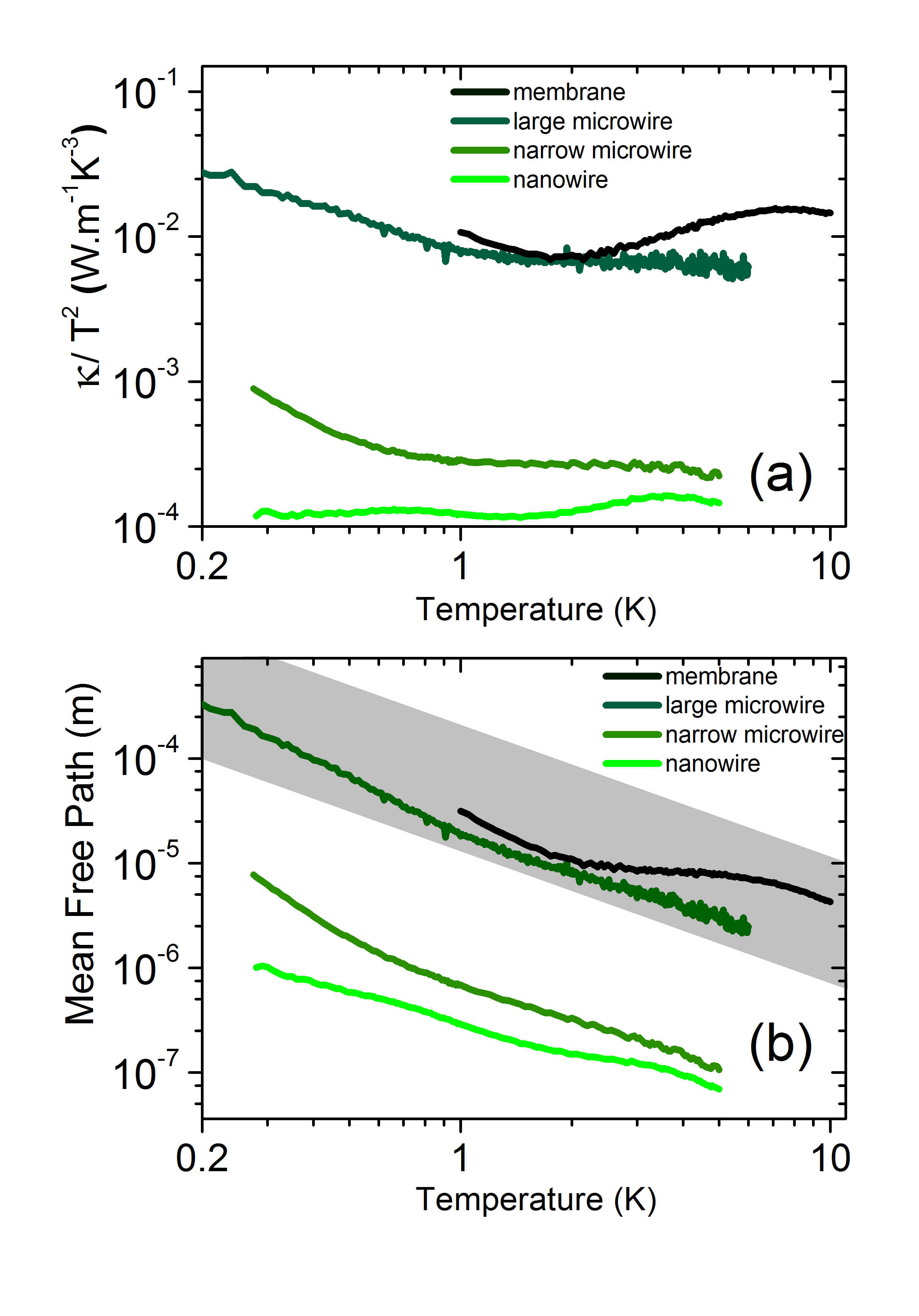}
 \end{center}
 \caption{(a) the thermal conductivity normalized to the square of the temperature ($\kappa/T^{2}$) in a log-log scale. (b) the phonon mean free path of the nanowire extracted from Eq.~\ref{MFPamorph}. The mean free paths are decreasing significantly as the size of the conductor is decreased, a clear signature of the impact of the low dimensions of the samples on the heat transport. The glassy range delimited by the grey area represents the maximum or minimum of mean free path measured in bulk amorphous materials (see ref.~\cite{RevPohl2002}).} 
 \label{fig3}
\end{figure}

In order to go deeper in the discussion and compare the dimensional reduction and TLS effects on phonon scattering on the thermal transport, one needs to report on thermal conductance normalized to length and widths \cite{note}. This is done by calculating the thermal conductivity $\kappa$ through the regular expression $\kappa=\frac{K L}{t w}$, where $K$ is the measured thermal conductance, $t$ the thickness of the materials and $w$ the width. This is shown in Fig.~\ref{fig2}~(b) where, for the same amorphous materials Si$_3$N$_4$, the thermal conductivities vary significantly from one geometry to the other, decreasing when the section of the heat conductor decreases.
This is illustrating the fact that the phonon transport is limited by boundary scatterings at low temperature, the so-called Casimir-Ziman regime of heat transport \cite{Casimir1939,Ziman}.

Deciphering the intrinsic mechanisms responsible for the heat transport implies obtaining the most relevant parameter: the phonon mean free path. In order to extract that crucial physical parameter from our measurements, one uses the phenomenological approach developed by Pohl, Liu and Thompson \cite{RevPohl2002} to interpret the thermal conductivity data obtained on bulk amorphous materials. The authors combine the well known kinetic relation $\kappa=\frac{1}{3}c_{D}v_{s} \Lambda_{ph}$ perfectly valid at low temperature (as long as ballistic transport is not involved) along with the fact that the thermal conductivity $\kappa$ is proportional to the square of the temperature as illustrated in Fig.~\ref{fig3}~(a):

\begin{equation}
\kappa(T)=\zeta_{0}T^2~=\frac{1}{3}c_{D}v_{s} \Lambda_{ph}
\label{kond1}
\end{equation}

where $\zeta_{0}$ is a phenomenological proportionality factor equal to the thermal conductivity $\kappa$ at 1~K; $c_{D}$ is the volumetric specific heat of the phonons carrying heat; $v_{s}$ is the average speed of sound, which is 9900~m/s in silicon nitride; and $\Lambda_{ph}$ is the phonon MPF independent of dimensions. $\zeta_{0}$ can be estimated through the TLS model, however this is not required for the determination of the phonon MFP, indeed $\zeta_{0}=\kappa_{T=1~{\rm K}}$, then directly extracted from the experiments.

One important fact should be clarified in this approach; one needs to know the specific heat $c_{D}$ related to phonons that are carrying heat. This should not be mixed up with the experimental specific heat of amorphous materials mostly dominated by the tunneling states \cite{RevPohl2002}. By using the Debye model for estimating $c_{D}$, very good agreement is obtained between experiments and calculations for temperatures down to 1~K as demonstrated by Pohl \textit{et al.} \cite{PohlPRB1971,Vuetal1998}. Similarly here we use the regular Debye expression at low temperature as the input for the specific heat:
\begin{equation}
c_{D}=\frac{2}{5}\frac{k_{B}^4}{\hbar^3}\frac{\pi^2}{v_{s}^3}T^3
\label{cDebye}
\end{equation}
When combining Eq.~\ref{kond1} and Eq.~\ref{cDebye} the phonon MFP can then be expressed by:
\begin{equation}
\Lambda_{ph}=\frac{15 \hbar^3 v_{s}^2}{2 \pi^2 k_{B}^{4}} \cdot \frac{\kappa(T) }{T^3}
\label{MFPamorph}
\end{equation}

One can see in Eq.~\ref{MFPamorph} that the MFP is given, not only by the temperature and the speed of sound, but also by the thermal conductivity as function of temperature. So, by measuring thermal conductivity one can have a direct experimental determination of the MFP as a function of the different sections of heat conductors. In our analysis, as shown in Fig.~\ref{fig3}~(a), we have first checked that the proportionality $\kappa \propto T^{2}$ is still valid even at the nanometer scale. Then, we have extracted the phonon MFP using Eq.~\ref{MFPamorph}; the results are presented in Fig.~\ref{fig3}~(b). 

Two different limits are observed for the phonon MFP depending on the size of the samples. The first concerns the large systems (membrane and large microwire) for which the MFP lies in the glassy limit given by the grey area Fig.~\ref{fig3}~(b). This glassy range is defined as the minimum or maximum MFP experimentally obtained on bulk amorphous materials. So concerning the large samples, we can conclude that the thermal transport is similar to what happens in bulk materials. On the other hand, for the small section samples, smaller phonon MFP's are clearly observed as if they were set by the interaction with the surfaces as expected in the Casimir-Ziman regime. The increase of MFP in narrow microwire and nanowire at low temperature is interpreted as the signature of specular reflections of phonons on the wire boundaries. When specular reflections are involved in the thermal transport, the temperature variation of the MFP is generally well described in the framework of the Berman Foster Ziman (BFZ) model of phonon boundary scattering using the sole physical roughness of the surfaces as a fitting parameter \cite{Berman}.

The phonon scattering on the boundaries may have two possible origins, either from the actual asperity (physical roughness) of the nanowires or due to the presence of TLS on the surface. We will then calculate an \textit{effective roughness} $\eta_{eff}$ of the nanowire from the experimental phonon MFP variation with temperature and compare it to the roughness estimated from the scanning electron microscopy image (SEM). That effective roughness will be representative of all the phonon scattering processes: scattering on boundaries characterized by the actual physical roughness (asperity) and the scattering of phonon on TLS.

To obtain the effective roughness for the nanostructures, we first extract the probability of specular reflection $p(\lambda,\eta)= exp\left(-16 \pi^3 \eta^2/ \lambda^2\right)$ where $\eta$ is the roughness of the nanowire's edges, and $ \lambda $ is the dominant phonon wavelength. To do that, we equal the experimental MFP to the MFP calculated from the BFZ model \cite{Berman}:
\begin{equation}
\Lambda_{ph}=\frac{1+p_{exp}(\lambda,\eta)}{1-p_{exp}(\lambda,\eta)} \Lambda_{Cas}
\label{MFPzim}
\end{equation} 

$\Lambda_{Cas}=1.12\sqrt{w\times t}$ is the Casimir MFP, where $ w\times t$ is the section of the nanosystems. So the experimental probability of specular reflection can be obtained from Eq.~\ref{MFPzim} through:
\begin{equation}
p_{exp}=\frac{\Lambda_{ph}-\Lambda_{Cas}}{\Lambda_{ph}+\Lambda_{Cas}}
\label{pzim}
\end{equation}
$\Lambda_{ph}$ is the experimental MFP as calculated through Eq.~\ref{MFPamorph} from the thermal conductance.

\begin{figure}
\begin{center}
 \includegraphics[width=10cm]{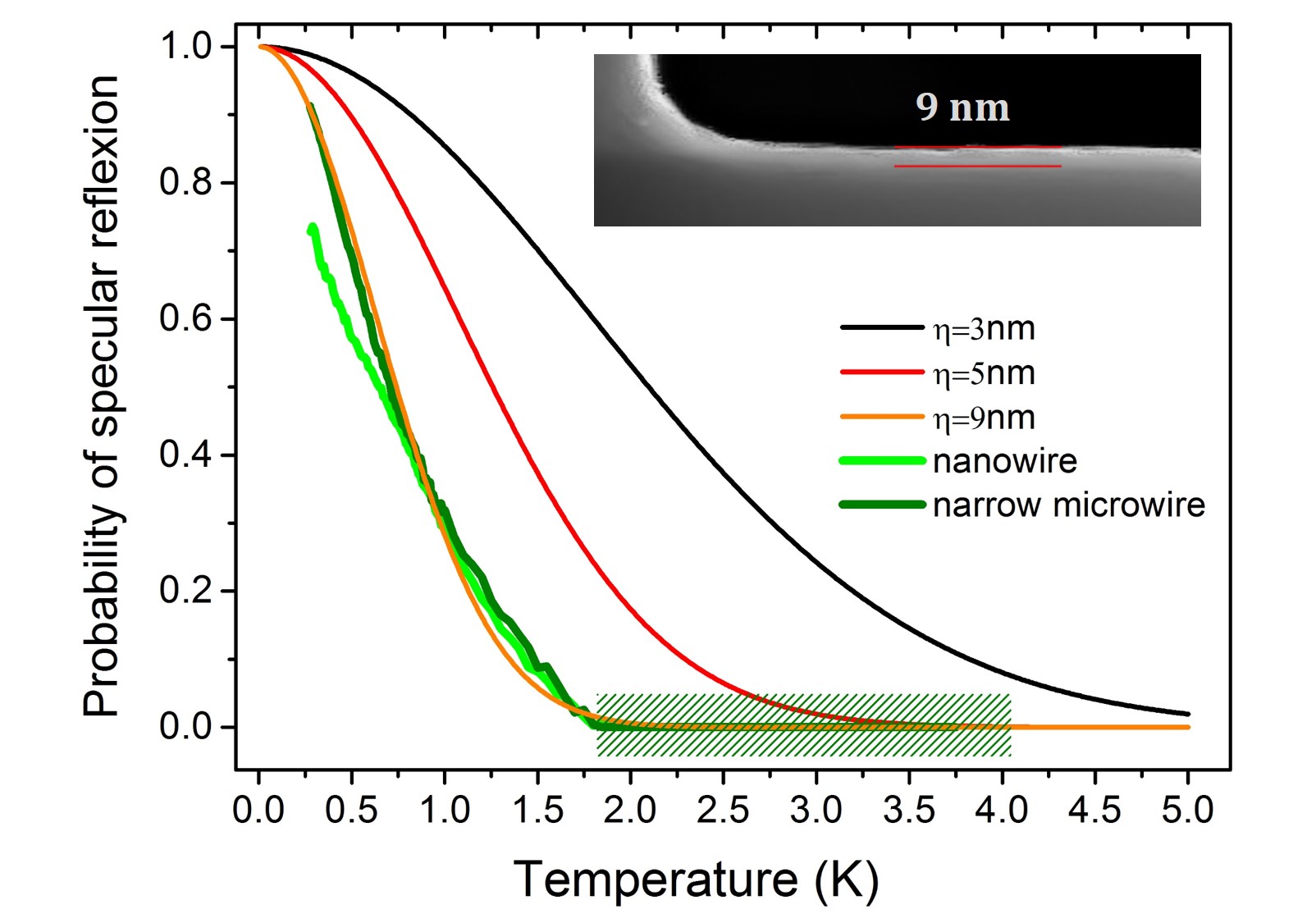}
 \end{center}
 \caption{Extracted probability of specular reflection for the nanowire (light green) and the narrow microwire (dark green) in comparison to the theoretical fit from Ziman approach using Eq.~\ref{pzim} with different roughness. The hatched area shows a purely diffusive regime known as the Casimir regime where the MFP becomes equal to the diameter of the nanostructure. In inset, an SEM picture of the SiN nanowire edge is shown. The actual roughness is much smaller than 9~nm.} 
 \label{fig4}
\end{figure}

The experimental probability of specular reflection for narrow microwire and nanowire is illustrated in Fig.~\ref{fig4} as extracted from Eq.~\ref{pzim}, in comparison with theoretical fits for different roughness. 
The fits help us to estimate this \textit{effective roughness} $\eta_{eff}$ that can be compared to the mean roughness obtained from the SEM characterization of the nanowire. The roughness that fits the experimental probability of specular reflection is of the order of $ \eta _{eff} \cong 9 $~nm three times bigger than the one evaluated by SEM observation which is about $ \eta\cong 3 \pm 1$~nm (see inset of Fig.~\ref{fig4}). This excess of roughness is attributed to TLS that act on the surfaces as an artificial roughness; meaning that phonon-surface TLS scattering dominates the heat transport. This is indeed fully consistent with the quadratic temperature variation of the thermal conductance \cite{note2}. 

To conclude, we show that for amorphous nanowires the temperature variation of the thermal conductance is still quadratic; even if it would have been expected that in restricted geometry the behavior of thermal conductance would be cubic like in the Casimir-Ziman regime (boundary scattering limit). This is ascribed to the presence of a strong density of phonon scattering centers located on the surface as seen in the study of the effective roughness obtained from the phonon mean free path. A possible high density of TLS can explain this observation which is in good agreement with the quadratic variation of $\kappa(T)$.
Actually, the TLS are expected to form in nanovoids or low density regions which are sensitive to preparation methods (temperature growth and thickness) of the materials \cite{queen2013,queen2015,liu2014,liupohl}. This means that, especially for thin films, the presence of voids in volume is less probable than on the surfaces. Consequently, the TLS may indeed be concentrated on the surfaces in agreement with our experimental observations \cite{queen2015}. 

Both results (quadraticity and phonon MFP) show the robustness of the universality of thermal transport in amorphous materials even down to the nanometer scale. The high density of TLS may have significant consequences for dissipation processes and decoherence phenomena in quantum nanoelectromechanical systems (NEMS) made out of amorphous SiN \cite{fefferman,defoort,maillet}.
Further experimental proofs of the high density of TLS on the surface could be obtained by specific heat measurements on low dimensional amorphous systems like very thin membranes at very low temperature or by nano-electromechanical measurement at very low temperature (below 10~mK). An abnormal high TLS density would be revealed by an anomalously high surface specific heat. 

We thank the micro and nanofabrication facilities of Institut N\'eel CNRS: the Pole Capteurs Thermom\'etriques et Calorim\'etrie (E. Andr\'e, P. Lachkar, G. Moiroux and J.-L. Garden) and Nanofab (T. Crozes, S. Dufresnes, B. Fernandez, T. Fournier, G. Juli\'e and J.-F. Motte) for their advices in the preparation of the samples. OB and EC acknowledges the financial support from the ANR project QNM Grant No. 040401, the European projects MicroKelvin EUFRP7 Grant No. 228464 and OB from MERGING Grant No. 309150.


\begin{thebibliography}{99}
\bibitem{note}
In the Casimir regime of heat transport, the calculated thermal conductivity $\kappa$ is meaningful only when comparing thermal transport between different sizes of sample of the same materials. Indeed, since the phonon MFP changes with the reduction of size in the case of nanostructured samples, the absolute value of thermal conductivity is not a relevant parameter.
\bibitem{note2}
We did not do the same data treatment to the two largest samples (microwire and membrane) because they belong to the quasi-bulk limit for which the Berman Foster Ziman model may not fully apply.
\end{thebibliography}
\end{document}